\documentclass{ws-procs9x6}
\newcommand{\lo}{leading order~}

\newcommand{\nlo}{next-to-leading order~}
\newcommand{\nloo}{next-to-leading order}

\def\Journal#1#2#3#4{{#1} {\bf #2}, #3 (#4)}  

\def\NPB{{\em Nucl. Phys.} B}  
\def\PLB{{\em Phys. Lett.}  B}  
\def\PRL{\em Phys. Rev. Lett.}

\begin{document}

\title{A Planned Jefferson Lab Experiment on Spin-Flavor Decomposition
}

\author{Xiaodong Jiang}

\address{Rutgers University, Piscataway, New Jersey. USA \\
E-mail: jiang@jlab.org}

\author{Peter Bosted,  Mark Jones }  
\address{Thomas Jefferson National Accelerator Facility,
 Newport News, Virginia, USA.} 

\author{Donal Day} 
\address{University of Virginia, Charlottesville, Virginia, USA.}

\maketitle

\abstracts{Experiment E04-113 at Jefferson Lab Hall C 
plans to measure the beam-target double-spin asymmetries in semi-inclusive
deep-inelastic $\vec p(e, e^\prime h)X$ and $\vec d(e, e^\prime h)X$ 
reactions ($h=\pi^+, \pi^-, K^+$ or$K^-$) with a 6 GeV polarized electron beam
and longitudinally polarized NH$_3$ and LiD targets.
The high statistic data will allow a spin-flavor 
decomposition 
in the region of
 $x=0.12 \sim 0.41$ at  $Q^2=1.21\sim 3.14$ GeV$^2$.
 Especially, \lo and next-to-leading order spin-flavor decomposition of 
$\Delta u_v$, $\Delta d_v$ and 
$\Delta \bar{u} - \Delta \bar{d}$ will be extracted based on 
 the measurement of the combined
 asymmetries $A_{1N}^{\pi^+ - \pi^-}$.  
The possible flavor asymmetry of the polarized sea will be addressed in this experiment. 
}

\section{Introduction}
Remarkable progress in the knowledge 
of the polarized quark distributions
  $\Delta q_f(x)$ has been made in the last decade through inclusive 
deep-inelastic lepton scattering (DIS). 
However, the information available from inclusive DIS process has inherent 
limitations.  As the cross 
sections are only sensitive to $e_q^2$, 
an inclusive  DIS experiment probes quarks and anti-quarks on an equal footing, therefore is not 
sensitive to the symmetry breaking in the sea sector. 
The sensitivity to each individual quark flavor can be  realized
in semi-inclusive deep inelastic scattering (SIDIS)
in which one of the leading hadrons is also detected.
Recently, the HERMES collaboration
published the results of a leading order spin-flavor decomposition from polarized 
proton and deuteron data, and for the first time
extracted the $\bar{u}, \bar{d}$ and $s=\bar{s}$ sea quark 
polarization~\cite{hermes2004}. 

The validity of the HERMES method of spin-flavor decomposition relies explicitly on 
several non-trivial assumptions. First, the leading-order ``naive $x$-$z$ factorization'' was assumed 
 and the next-to-leading order terms were neglected.
 This implies that the cross sections
 factorize into  the $x$-dependent quark distributions and the $z$-dependent 
 fragmentation functions: 
\begin{eqnarray} 
 \sigma^{h} (x,z) & = & \sum_{i} e_f^2 q_f(x) \cdot D_{q_f}^{h}(z),  \hspace{0.15cm}
 \Delta \sigma^{h} (x,z)  =  \sum_{i} e_f^2 \Delta q_f(x) \cdot D_{q_f}^{h}(z).  
\label{eq:fact}  
\end{eqnarray}  
Furthermore, it was assumed that the quark fragmentation process and the experimental 
phase spaces were well-understood such that a LUND model 
based Monte Carlo simulation program can 
reliably reproduce the ``purity matrices'' which account for the probability correlations
between the detected hadrons and the struck quarks~\cite{hermes2004}.

 It was pointed out by Christova and Leader~\cite{leader2} that 
if the combined asymmetries $A_{1N}^{\pi^+ - \pi^-}$ are measured,
quark polarization $\Delta u_v$, $\Delta d_v$ and $\Delta \bar{u} - \Delta \bar{d}$ can be extracted
at \lo without the complication of fragmentation functions.  Even 
at \nloo, information on the 
valence quark polarizations is well preserved in the combined asymmetries $A_{1N}^{\pi^+ - \pi^-}$.

\section{The Christova-Leader method at LO and NLO}
At the leading order, 
under isospin symmetry and charge conjugation, the fragmentation functions 
cancel exactly in the combined asymmetry $A_{1N}^{\pi^+ - \pi^-}$, 
the $s$-quark does not contribute, and we have~\cite{leader2} :
\begin{eqnarray}
\label{Eq:cl1}
&A&\hspace{-0.1cm}_{1p}^{\pi^+ - \pi^-}  =  { \Delta \sigma_p^{\pi^+}-\Delta \sigma_p^{\pi^-} \over
\sigma_p^{\pi^+} - \sigma_p^{\pi^-} }=
{  4\Delta u_v - \Delta d_v 
\over 4u_v - d_v },  \nonumber \\
&A&\hspace{-0.1cm}_{1d}^{\pi^+ - \pi^-}  =  { \Delta \sigma_d^{\pi^+}-\Delta \sigma_d^{\pi^-} \over
\sigma_d^{\pi^+}- \sigma_d^{\pi^-} }=
{ \Delta u_v + \Delta d_v 
\over u_v + d_v}.
\end{eqnarray}
Therefore, measurements of $A_{1N}^{\pi^+ - \pi^-}$ on the proton and the 
 deuteron can determine $\Delta u_v$ and $\Delta d_v$. 
On the other hand, the existing inclusive DIS data already constrains another non-singlet quantity:
\begin{equation}
\label{Eq:nlog1pn}
g_1^p(x,Q^2) - g_1^n(x,Q^2) = { 1 \over 6 } \left[ (\Delta u + \Delta \bar{u}) - (\Delta d + \Delta \bar{d}) \right] 
\vert_{LO}.
\end{equation}
The polarized sea asymmetry can be extracted at \lo following:
\begin{equation}
(\Delta \bar{u} - \Delta \bar{d}) \vert_{LO} = 3 (g_1^p- g_1^n)\vert_{LO} 
 - {1 \over 2} (\Delta u_v - \Delta d_v) \vert_{LO}.
\end{equation}

At the next-to-leading order, $x$ and $z$ are mixed through double convolutions, 
 and instead of Eq.~\ref{eq:fact}, we have:
 \begin{eqnarray}
\label{Eq:nlo1}
\sigma^h(x,z) & = & \sum_{f}  e_f^2 q_f \left[ 1 + \otimes {\alpha_s \over 2 \pi} {C}_{qq} \otimes \right] 
          D_{q_f}^{h}  \nonumber \\ 
& + & \left( \sum_{f}  e_f^2 q_f \right) \otimes {\alpha_s \over 2 \pi} {C}_{qg} \otimes D_G^h
+ G \otimes {\alpha_s \over 2 \pi} {C}_{gq} \otimes \left( \sum_{f} e_f^2 D_{q_f}^{h} \right)  
\end{eqnarray}
and similarly for $\Delta \sigma^h$, where $C$s are well-known Wilson coefficients.
 The convolution terms become much simpler~\cite{leader2} in quantities  
relate to $\sigma^{\pi^+} - \sigma^{\pi^-}$ since the $gq$ and $qg$ terms in 
 Eq.~\ref{Eq:nlo1} are identical for $\pi^+$ and $\pi^-$: 
 \begin{eqnarray}
\label{Eq:a1pnlo}
A_{1p}^{\pi^+ - \pi^-} & = &  { (4 \Delta u_v -\Delta d_v) \left[ 1+ \otimes (\alpha_s/2\pi) \Delta C_{qq}
\otimes \right] (D^+ -D^-) \over { (4 u_v - d_v) \left[ 1+ \otimes (\alpha_s/2\pi) {C}_{qq}
\otimes \right] (D^+ - D^-) } }, \nonumber \\ 
A_{1d}^{\pi^+ - \pi^-} & = &  { (\Delta u_v + \Delta d_v) \left[ 1+ \otimes (\alpha_s/2\pi) \Delta C_{qq}
\otimes \right] (D^+ -D^-)  \over { (u_v + d_v) \left[ 1+ \otimes (\alpha_s/2\pi) {C}_{qq}
\otimes \right] (D^+ -D^-) } }.
\end{eqnarray}
in which $\Delta u_v$ and $\Delta d_v$ evolve as non-singlets and do not mix with
gluon and sea distributions. Once we extract $\Delta u_v$ and $\Delta d_v$ at \nlo from Eq.~\ref{Eq:a1pnlo}, 
 $\Delta \bar{u} - \Delta \bar{d}$ can be determind to \nlo using
the well-known NLO form of Eq.~\ref{Eq:nlog1pn}. 

\section{The Jefferson Lab experiment E04-113}

 Experiment E04-113 at Jefferson Lab Hall C~\cite{e04113} is specifically designed to have well controlled phase spaces
 and hadron detection efficiencies such that the combined asymmetries $A_{1N}^{\pi^+ \pm \pi^-}$, 
 in addition to the individual asymmetries $A_{1N}^h$ 
($h=\pi^+, \pi^-,K^+,K^-$), can be determined
with high precision.
The existing HMS spectrometer will be used as the hadron detector at $10.8^\circ$ 
and a central momentum of 2.71 GeV/c, corresponding to $\langle z \rangle \approx 0.5$ to favor the current fragmentation.
For the electron detector, a combination of a large calorimeter array and a gas Cherenkov 
will be used. 
The experiment will cover $0.12<x<0.41$ with $1.21<Q^2<3.14$ (GeV/c)$^2$ and $2.31<W<3.09$ GeV. 

\begin{figure}[htbp]
\centerline{\epsfxsize=2.5in\epsfbox{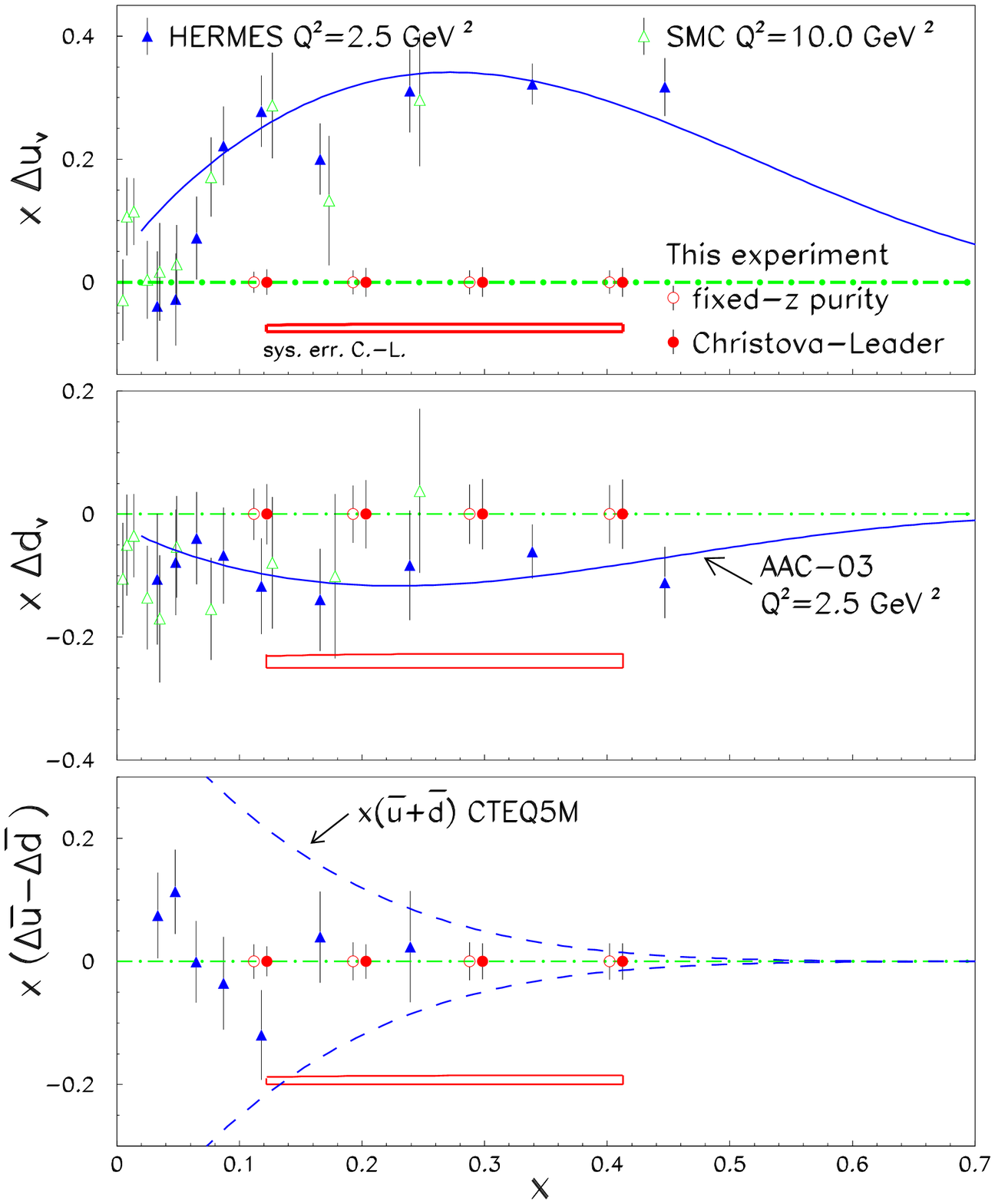} \epsfxsize=2.3in\epsfbox{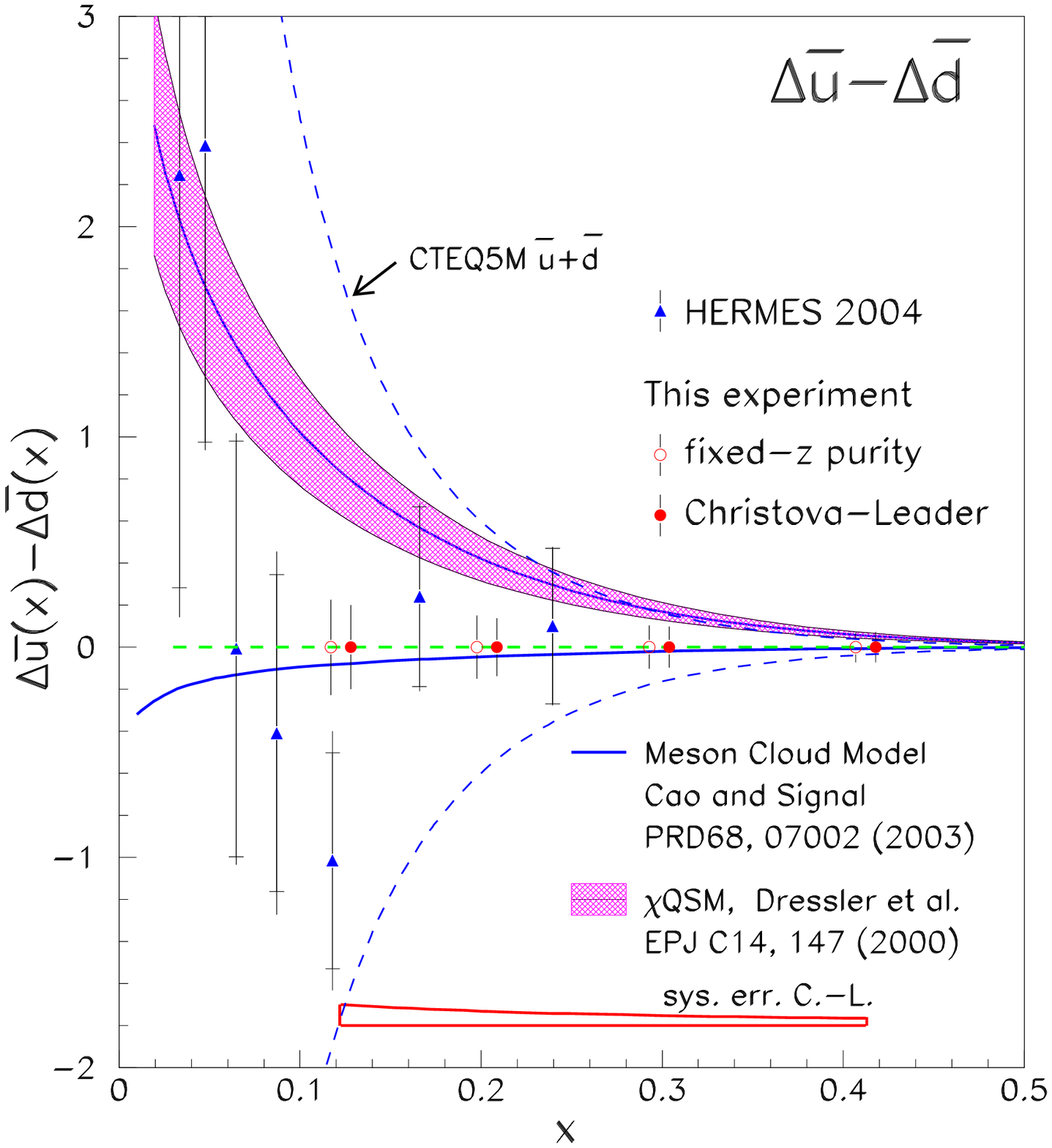}}   
\caption{
\label{fig:xdelqv} The expected statistical accuracies of experiment E04-113 for two independent methods 
of flavor decomposition (Christova-Leader and ``fixed-$z$ purity'') 
are compared with the HERMES data~\protect\cite{hermes2004} and the SMC data~\protect\cite{smc1998}.  
The open boxes represent the systematic uncertainties of the Christova-Leader method.
}
\end{figure}

In addition to the Christova-Leader method, the ``fixed-$z$ purity'' method of spin-flavor decomposition 
will be applied to provide a consistency check. At the well-defined $z$-value of this experiment, 
the ``purity matrices'' can be directly calculated 
based on unpolarized PDFs and the ratio of fragmentation functions, rather than from a Monte Carlo simulation which
involves a fragmentation model. The expected statistical accuracies and the estimated systematic
uncertainties are shown in Fig.\ref{fig:xdelqv}.

\section{Conclusions}
 Experiment E04-113 at Jefferson Lab plans to extract quark polarizations
  based on the measurement of the combined asymmetries $A_{1N}^{\pi^+ - \pi^-}$. The much improved statistics
 over the HERMES data will present us with the first opportunity 
 to probe the possible flavor asymmetry of the light sea quark 
 polarization. 

We thank Drs. E. Christova, E. Leader, G.A. Navarro, R. Sassot, D.~de~Florian, A.~Afanasev, W.~Melnitchouk
for many discussions. This work is supported in part by the 
US Department of Energy and the US National Science Foundation.

\end{document}